%% file: arxiv.tex
\documentclass[11 pt, a4paper]{article}

\usepackage{array,epsfig}
\usepackage{color}
\usepackage{enumitem}
\usepackage{sidecap}
\usepackage{setspace}
\usepackage{comment}
\usepackage{url}
\usepackage{float}
\usepackage{cite}
\usepackage{apacite}
\usepackage{parskip}
\usepackage{standalone}
\usepackage{tabu}
\usepackage{longtable, booktabs}
\usepackage{lscape}
\usepackage{xltabular}
\usepackage[font=footnotesize,labelfont=bf]{caption}
\usepackage{bm, bbm, mdframed, mathtools, mathabx, multirow, subfigure, tikz, multirow, tcolorbox, soul, caption}
\usepackage[left=2.5cm, right=2.5cm, top=3cm, bottom=3cm]{geometry}
\usepackage[english]{babel}
\usepackage[utf8]{inputenc}
\usepackage[mathscr]{euscript}

\numberwithin{equation}{section} 
\tcbuselibrary{breakable}
\usepackage{mathdots}
\usepackage{lscape}
\usepackage{pdflscape}

\usepackage{svg}
\usepackage[round]{natbib}
\usepackage{hyperref}
\hypersetup{%
  colorlinks = true,
  citecolor = black,
  filecolor=black,
  urlcolor=black,
  linkcolor = black
}
\usepackage{glossaries}
\usepackage{appendix}
\usepackage{verbatim}
\usepackage{tcolorbox}
\usepackage{xcolor}
\usepackage{listings}
\usepackage{rotating}
\usepackage{comment}
\usepackage{graphicx}
\usepackage{dsfont}
\usepackage{adjustbox}
\usepackage[para,online,flushleft]{threeparttable}
\usepackage{ragged2e}
\usepackage{multirow}
\usepackage{tabularx}

\usepackage{amsmath}
\usepackage{algorithm}
\usepackage{algpseudocode}

\title{\textbf{Enabling Decision-Making with the Modified Causal Forest: Policy Trees for Treatment Assignment}}

\author{
  Hugo Bodory\thanks{University of St.Gallen, Varnbuelstrasse 14, 9000 St.Gallen, CH, E-mail:\texttt{hugo.bodory@unisg.ch}, \texttt{federica.mascolo@unisg.ch},  \texttt{michael.lechner@unisg.ch}\\
   } , Federica Mascolo \textsuperscript{\footnotemark[1]},
    Michael Lechner\textsuperscript{\footnotemark[1]}
\textsuperscript{\footnotemark[2]}\thanks{Michael Lechner is also affiliated with CEPR, London, CESIfo, Munich, IAB, Nuremberg and IZA, Bonn, RWI, Essen. \\
Financial support from the Swiss National Science Foundation (SNSF) is gratefully acknowledged. The study is part of the project "Chances and risks of data-driven decision making for labour market policy" (grant number SNSF 407740\_187301) of the Swiss National Research Programme "Digital Transformation" (NRP 77). We thank GPT-3.5 and Grammarly for their editorial help.}
}

\begin{document}
\begingroup
\let\newpage\relax
\maketitle
\endgroup
\begin{center}
 \vspace{1.2cm}
  \textbf{Abstract}   
\end{center}
\begin{minipage}{\textwidth}
    \small
Decision-making plays a pivotal role in shaping outcomes in various disciplines, such as medicine, economics, and business. This paper provides guidance to practitioners on how to implement a decision tree designed to address treatment assignment policies using an interpretable and non-parametric algorithm. Our Policy Tree is motivated on the method proposed by \cite*{zhou2023offline}, distinguishing itself for the policy score calculation, incorporating constraints, and handling categorical and continuous variables. We demonstrate the usage of the Policy Tree for multiple, discrete treatments on data sets from different fields. The Policy Tree is available in Python's open-source package \textit{mcf} (Modified Causal Forest).

\vspace{10mm}

\textbf{JEL classification:} C21; C87; C14; C69 \\
\textbf{Keywords:} causal machine learning; statistical learning; conditional average treatment effects; individualised treatment effects; multiple treatments; optimal policy; offline policy learning; decision trees
\end{minipage}

\thispagestyle{empty}



\section{Introduction}
Policymakers are keen to gain insights into policy impacts, aiming to facilitate future applications. Therefore, current econometrics research is dedicated to developing methods that accurately estimate policy effects, specifically focusing on understanding relevant heterogeneity across sub-populations. However, it is important to distinguish between estimating heterogeneity and providing guidance on the design of treatment assignment mechanisms. The latter pertains to the policy learning research area, which effectively uses information from randomised or observational studies to assign individuals to future treatments. 

The seminal work of \cite*{manski2004statistical} pioneered the literature on policy learning, then further developed by \cite*{kitagawa2018should, mbakop2021model, athey2021policy, kitagawa2021equality} and \cite{zhou2023offline}, among others. Specifically, \cite*{kitagawa2018should} introduce a policy learner that attains the $\sqrt{N}$ convergence rates in settings with a binary treatment and known propensity scores. \citet{athey2021policy} propose a policy learner that achieves the same convergence rate even without known propensity scores, making it well suited for observational studies. Finally, \cite{zhou2023offline} propose the Cross-fitted Augmented Inverse Propensity Weighted Learning (CAIPWL) policy learner, which achieves optimal convergence rates in multi-action settings with unknown propensity scores.  
 
This work presents the \textit{mcf} Policy Tree as implemented in the \textit{mcf} package. The policy learner is based on the algorithm proposed by \cite{zhou2023offline} but introduces three key modifications: policy scores calculation, implementation of constraints, and different handling of categorical and continuous variables. First, the consistent and asymptotically normally distributed policy scores (or potential outcomes) are computed by the \textit{mcf} to calculate individualised average treatment effects (IATEs)\citep*{lechner2024comprehensive}. They differ from the doubly-robust scores used in the original algorithm since the mcf splitting criterion differs. In particular, the \textit{mcf} algorithm minimises the estimation errors of the conditional mean responses of the treatments and the covariance of these errors while also taking into account propensity score heterogeneity to reduce selection bias. Additionally, the  IATEs differ from doubly robust scores since they depend only on the features and not the outcome and treatment.
The second modification concerns the implementation of constraints. They are integrated as treatment-specific cost values in terms of the outcome, which are then subtracted from the policy scores. This also applies to constraints that are only indirectly related to costs, such as constraints on treatment shares. 
Finally, the \textit{mcf} Policy Tree differs from existing methods in the literature in handling categorical and continuous features. Traditional approaches in the literature often rely on one-hot encoding for categorical features, which may lead to rather extreme leaves with many different values when building shallow trees due to splitting one category against the remaining ones. In contrast, the \textit{mcf} Policy Tree employs a distinct procedure that allows several categories on both sides of a split. For a detailed explanation of this approach, refer to Chapter 9.2.4 in \cite*{hastie2009elements}.
Beyond this functionality, the \textit{mcf} algorithm provides flexibility for handling continuous and categorical, unordered features with many possible values by allowing users to specify an approximation parameter. This parameter determines the number of potential splitting points, offering finer granularity\footnote{4th level, i.e., at the top of the tree, A/8; 3rd level A/4; 2nd level A/2; and the 1st level, at the bottom of the tree, 
A; with A being the approximation parameter as used in \cite{zhou2023offline} as one single global approximation level.} compared to the global approximation parameter implemented in \cite{zhou2023offline}.

This paper contributes to the literature in various ways. First, we demonstrate the application of the Policy Tree algorithm across diverse fields, including medicine, epidemiology, and business, within both randomised and observational research settings. Second, this work provides new empirical results on causal effect heterogeneity and policy learning. In that respect, we evaluate well-known and used data in the respective scientific fields. Third, our results take cost-related constraints for decision-making into account. Fourth, all replication materials, including data and programming codes, can be accessed on Github. To run these codes, the open-source Python package \textit{mcf} provides the classes \texttt{ModifiedCausalForest()} for estimation and \texttt{OptimalPolicy()} for decision-making. 

In the remainder of this paper, Section \ref{Framework} introduces the framework for identification and estimation and package functionalities. Section \ref{ES} shows the setups and findings of the empirical studies. Finally, Section \ref{DC} discusses computational facets and concludes.

\section{Framework}\label{Framework}

The \textit{mcf} package is a forest-based causal machine learning algorithm that produces consistent and asymptotically normal treatment effect estimates for various levels of granularity in randomised control trials and selection-on-observable settings. The latter imposes several assumptions to identify the causal effects, namely ignorability, overlap, the stable unit treatment value assumption, and exogeneity of features; see \citet{ImWo2009} for more details on these identifying assumptions. 
The estimands of interests identify the causal effect at different aggregation levels, in particular, the average potential outcomes, the average treatment effect, and conditional average effects like IATEs and grouped aggregates of these effects. The IATEs are the average effects for the most granular sub-groups, identified by specific realisations of the available exogenous features. The evaluation of the IATEs makes it possible to detect potential heterogeneous effects across sub-groups of the population. If heterogeneity is observed, certain individuals may either benefit or not from a particular treatment. This is a crucial starting point for a policy learning analysis, indicating the potential for refining the treatment assignment mechanism and informing policy recommendations.

To address this, we introduce the \texttt{OptimalPolicy()} class of the \textit{mcf}, which explores the space of all available trees and chooses the optimal one. The optimal tree maximises the value function (or welfare), computed as the sum of the individual policy scores, such as potential outcomes or IATEs, by assigning all observations in a terminal leaf node to a single treatment. For a fixed depth, the tree algorithm implements an exhaustive and recursive search across all pre-specified variables and values. It performs the splits when the policy scores within the leaves (or subsets) are maximised under a specific treatment. 

Algorithm \ref{algo:policy_tree} illustrates, in a simplified way, how the tree-search algorithm works. To introduce it, we can assume the following notation:
For \( i \) observations, ${X_i}$ represents the features, specifically, \( p_1 \) is the number of ordered features, \( p_2 \) is the number of unordered features, $j\in\{0,\ldots, D \}$ is the treatments space, \( \hat{\Theta}_i \) is the vector (or matrix) of policy scores for each observation \( i \), \( \hat{\Theta}_i(j) \) is the potential outcome for each observation for each treatment \( j \), and \( L \) is an integer indicating the depth of the tree plus one.\footnote{For detailed reference of the algorithm \href{https://github.com/MCFpy/mcf/blob/ae13663d0a8e86c7d123471bd7741489a39401d6/mcf/optpolicy_pt_eff_functions.py}{optpolicy\_pt\_eff\_functions.py}.}

\begin{algorithm} 
\caption{Policy Tree} 
\label{algo:policy_tree}
\begin{algorithmic}[1]
    \State \textbf{Input:} $\{(X_i, \hat{\Theta}_i(j))\}_{i=1}^n$, $L$
    \If{$L = 1$}
        \State \textbf{Return} $(\max _{j \in\{1, \ldots, d\}} \sum_{i}\hat{\Theta}_i(j), \operatorname{argmax}_{j \in\{1, \ldots, d\}} \sum_{i}\hat{\Theta}_i(j))$
    \Else
        \State Initialise reward $\mathcal{R} \leftarrow -\infty$, and empty tree $\mathcal{T} \leftarrow \emptyset$ for all $m = 1,\ldots,p_1 + p_2$
        \For{$m \leftarrow 1, 2, \ldots, p_1 + p_2$} 
            \For{sorted values of ordered or unique categories of continuous m-th features}
                \State $\text{reward\_left}, \text{tree\_left} \leftarrow \text{Tree-Search}(set_{L}, L - 1)$
                \State $\text{reward\_right}, \text{tree\_right} \leftarrow \text{Tree-Search}(set_{T}, L - 1)$
                \If{$\text{reward\_left} + \text{reward\_right} > \mathcal{R}$}
                    \State $\mathcal{R} \leftarrow \text{reward\_left} + \text{reward\_right}$
                    \State $\mathcal{T} \leftarrow \text{Tree-search}(m, \text{splitting value}, \text{tree\_left}, \text{tree\_right})$
                \EndIf
            \EndFor
        \EndFor
        \State \textbf{Return} $(\mathcal{R}, \mathcal{T})$
    \EndIf
\end{algorithmic}
\end{algorithm}

Besides the Policy Tree, the \texttt{OptimalPolicy()} class provides a \texttt{best-score} method to create an assignment rule. This method simply assigns units to the treatment with the highest estimated potential outcome. Although computationally cheap, this \texttt{best-score} method lacks clear interpretability for the allocation rules, which may make it difficult for policymakers to adopt it.

\section{Empirical Studies}\label{ES}

The \textit{mcf} package offers empirical researchers dual valuable functionalities: the possibility of conducting heterogeneity analyses, as shown in \cite*{bodory2022high}, and refining the decision-making process by optimising treatment allocation rules. 
This section studies three real-world applications from different scientific fields based on distinct research settings. In these empirical tasks, we first perform an effect heterogeneity evaluation calling the \texttt{ModifiedCausalForest()} class of the \textit{mcf} algorithm. We then use the potential outcomes estimated by the \textit{mcf} to optimise the treatment allocations generated by the Policy Trees using the \textit{mcf}'s \texttt{OptimialPolicy()} class. The samples from all three studies are split using the same approach. Specifically, 40\% of the data is allocated for training the \textit{mcf}, and another 40\% for estimating the effects and training the policy learning algorithm. Finally, the remaining 20\% of the data are reserved for predicting out-of-sample policy rules and welfare. This allocation helps to estimate the treatment effect effectively and to mitigate the risk of overfitting, particularly for deeper trees.

\subsection{Study 1: Oregon Health Insurance Experiment}\label{OHE}

In 2008, Oregon initiated a lottery with limited spots to provide low-income individuals access to its Medicaid program. This lottery offers an opportunity to examine the impact of expanding access to public health insurance within a randomised controlled setting. Researchers analysed various outcomes over two years following the experiment, including healthcare utilisation, financial strain, health status, labour market outcomes, and political participation. Extensive studies have been conducted by \cite{finkelstein2012oregon, baicker2013oregon, finkelstein2016effect, baicker2017effect,/content/paper/687686456188, baicker2018impact, finkelstein2019value}, and \cite*{baicker2014impact}, among others. Overall findings suggest that access to Medicaid leads to an increase in healthcare utilisation and a decrease in financial strain and depression, albeit without statistically significant improvements in physical health or labour market outcomes. 
This analysis follows \cite{finkelstein2012oregon} and specifically focuses on the impact of being selected in the lottery on primary care visits during the first year post-selection. We rely on survey data collected within the experimental setting,  given that the hospital records are not publicly available. The end-line survey, conducted after 12 months, was completed by 23,777 individuals. The outcome is measured by a binary variable indicating if primary care has been utilised during the year. Among the several outcomes available (e.g. emergency room usage or drug prescription), we choose primary care utilisation because of its importance from a policy perspective. Indeed, primary care visits may lead to preventive care for more serious diseases and, consequently, prevention of more expensive medical care.

After dropping observations with missing values, the size of our dataset is 23,527 observations, with  50\% of the individuals with Medicaid health insurance. As in \cite{finkelstein2012oregon}, our intention-to-treat specification incorporates additional features, such as socio-demographic factors, which are valuable for exploring heterogeneity and informing policy, even though they are not essential for identifying causal effects. On the contrary, although assignment to the lottery is random, all family members of randomly selected individuals can apply for Medicaid. Therefore, the probability of receiving treatment depends on the number of family members, which we include as a control variable. Table \ref{varlist_ohe} in the Appendix describes the 11 variables used for data analysis. 

\input{OHE/iates_dens_ohe}

We evaluate the distribution of the IATEs to detect potential effect heterogeneity.  Figure~\ref{fig:iates_densities} displays a density plot of the IATEs, ranging from -0.13 to 0.37, with a mean value of 0.07 and an average standard error of 0.1. The variation in the effects justifies our following analysis for decision-making, aiming at expanding the number of primary care visits by optimising the treatment allocation process. 

Table \ref{table:policy_tree_ohe} provides details on welfare, measured as utilisation of primary care visits and allocation shares of individuals for various policies. We compare the results of the Policy Trees with our two baseline policies, namely the observed and random treatment allocations. In this study, we examine both unconstrained and constrained Policy Trees. Given the lack of specific cost information for Medicaid insurance,
we assume that only approximately half of the population (as in the empirical data) can be covered by this health insurance.
The algorithm enforces constraints by finding a cost value to subtract from the policy score. It adjusts the treatment costs to ensure that the allocation of observations to treatments does not exceed the maximum treatment shares. We build our constrained Policy Trees by specifying either the depth of a first optimal tree (Constrained optimal Policy Tree) or, optionally, specifying, in addition, the depth of a second optimal policy tree  (Constrained sequentially optimal Policy tree). In the latter case, the second tree is built within the strata obtained from the leaves of the first tree. The final Policy Tree will not be optimal. However, its performance in terms of welfare is comparable to that of an optimal tree with the same depth, with the advantage of reduced computational time.\footnote{The sequentially optimal policy tree is implemented as the \textit{hybrid policy tree} within the \texttt{policytree} package (\cite{policytree})}

\input{OHE/policy_out_ohe}

In Table~\ref{table:policy_tree_ohe}, all the Policy Trees report an increase in welfare with respect to the baseline policies. The unconstrained Policy Trees achieve the highest welfare, increasing the probability of primary care visits from 51.70\% to 54.98\%. They allocate almost all individuals to the treatment, which can be interpreted as access to Medicaid insurance for nearly everyone. Compared to the unconstrained Policy Trees, those with the capacity constraint always report a small decrease in the probability of primary care visits, allocating a significantly lower proportion of individuals to the treatment. Constrained optimal Policy Trees and Constrained sequentially optimal Policy Trees, with a last sequential optimal sub-tree of the same depth, show comparable and stable results in terms of welfare achieved. From a computational point of view, such findings may be relevant for large datasets since Policy Trees with the last sequential optimal layers reduce computation time by performing similarly well.

Figure \ref{fig:policy_rules_ohe} presents the allocation rules of a constrained optimal Policy Tree with a depth of 2. Notably, even at this relatively shallow depth, the Policy Tree demonstrates improvements in welfare compared to the observed allocations and the simplest allocation rules.  Individuals with age between 37 and 51 years old (38\% of the sample) are assigned to Medicaid. Additionally, individuals older than 51 living in metropolitan areas, constituting 9\% of our sample, are also allocated to the program. Conversely, individuals younger than 36 and older than 51 who do not live in metropolitan areas are excluded from the program. For reference, note that the age ranges from 20 to 63.

\input{OHE/policy_tree_ohe}

\subsection{Study 2: Right Heart Catheterization} \label{RHC}

In a seminal study, \cite*{connors1996effectiveness} investigated the impact of Right Heart Catheterization (RHC) on various outcomes, including subsequent survival, healthcare costs, intensity of care, and length of hospital stay. Their findings revealed a positive association between RHC and mortality, as well as increased costs and prolonged length of stay. This conclusion was corroborated by subsequent research utilising different estimation methods, such as those employed by \citet{knaus1995support}, \citet{ramsahai2011extending}, \citet{keele2018pre}, and \citet{keele2021comparing}. In accordance with previous research, we also use the dataset originally collected from the SUPPORT prospective cohort study (Study to Understand Prognoses and Preferences for Outcomes and Risks of Treatments) as in \cite{connors1996effectiveness}.
This dataset includes 5,735 observations on ill and hospitalised individuals between 1989 and 1994 in five medical centres in the US. 38\% of the patients received the surgery. \cite{bodory2022high} utilise the same information to provide new results on effect heterogeneity by exploring the functionalities of the \textit{mcf} package.

In this empirical application, we focus on finding an optimal allocation of RHC to patients to increase their probability of survival six months after the medical surgery. Previous work exploits expert opinion to identify eight features as relevant confounding factors \citep{ramsahai2011extending}. These features include an index of activities of daily living two weeks before admission, age, the acute physiology and chronic health evaluation score (APACHE III), the Glasgow coma score, mean blood pressure, the probability of surviving two months (based on model estimations), an indicator for do not resuscitation status on the first day, and information on nine primary disease categories. Recognising the significance of these factors, we employ them as features in the Policy Tree to obtain assignment rules for decision-making.
Age and mean blood pressure are continuous variables, and we allow the tree algorithm to automatically determine the optimal split points based on the default number of evaluation points in the \textit{mcf} package.

Our analysis starts by estimating the policy scores and IATEs. Additionally to the high-priority factors, we include potential confounding features (e.g. information on having cancer, possessing health insurance, etc.) to achieve identification through a selection-on-observable approach. Table \ref{varlist_rhc} of the Appendix comprehensively describes all 55 confounders. 

\input{RHC/IATE/iates_density}

Figure \ref{fig:iates_dens_rhc} displays the density function of the IATEs, the most granular treatment effects. While slightly right-skewed, the distribution shows positive and negative treatment effects, ranging from -0.12 to 0.08, and an average standard error of 0.06. This result underscores the (rather moderate) heterogeneous response to the surgery, indicating that while some individuals benefit from the intervention, others do not. Such heterogeneity emphasises the importance of conducting a treatment allocation analysis to investigate potential improvements in the allocations of individuals.

Table \ref{table:policy_tree_rhc} shows that simple allocation schemes may improve the overall survival rate while decreasing the number of surgeries at the same time. Again, the first two rows of Table \ref{table:policy_tree_rhc} present results using the empirically observed and random assignment of individuals to treatments, which are used as baseline comparison policies. The remaining rows show the results of Policy Trees aiming to optimise the treatment allocation and increase the survival rate.
Given the relatively small sample size, this study focuses on optimal trees of depths two and three, alongside Policy Trees with depths 3 and 4, including a final sequential optimal sub-tree (+1).

\input{RHC/policy_tables/policy_out}

Table \ref{table:policy_tree_rhc} first reveals that the observed treatment allocation outperforms a random treatment assignment in terms of the survival rate (68.23\% vs. 67.88\%). Considering the performance of the Policy Trees, the findings suggest treatment allocations that could further increase the average survival rate within six months after the intervention up to 69.99\% while significantly diminishing the number of individuals receiving surgery by more than 50\%. The maximum welfare gain, among the Policy Trees estimated, is already achieved with a Policy Tree of depth two. Deeper trees do not lead to a relevant increase in the survival rate. However, they reduce the share of individuals undergoing an operation. For example, compared to the depth-2 tree, the depth-3+1 tree indicates a reduction in survival rate by only 0.02 percentage points (69.99\% vs. 69.97\%) but decreases the share of patients having surgery by nearly three percentage points (19.21\% vs. 16.22\%). 

Since the maximum welfare gain is already attained with a depth-2 tree,  Figure \ref{fig:tree_alloc_rules} displays the decision rules established by this tree, which is also the most straightforward to interpret.
The Policy Tree performs the first split on age, while the second splitting variable is the estimated probability of surviving within two months, measured before the surgery. The decision tree allocates individuals to surgery with an age below 65 years and a survival probability below 48\%, as well as with an age above 65 years and a survival probability below 40\%. Note that tables representing all Policy Tree rules used are available in the Appendix.

\input{RHC/policy_tables/rhc_tree}

\subsection{Study 3: Interest rates and loan returns} \label{CE}

In 2003, \cite*{karlan2008credit} conducted a randomised controlled trial to examine the impact of personalised pricing by a microfinance lender in South Africa. They randomised interest rates to study the effect on the decision to borrow (extensive margin) and, among other outcomes, on the lender's revenues (intensive margin). Considering the intensive margin, the overall findings suggest that lower interest rates only marginally decrease profits. 

This study is particularly relevant because microfinance initiatives may serve broader purposes beyond profit maximisation. They can play a crucial role in promoting financial inclusion among specific population subgroups. Moreover, any initial losses incurred by the lender could be offset by a wider uptake of microfinance services.

Therefore, this study employs the same data and the intention-to-treat strategy of \cite{karlan2008credit} to investigate the impact of receiving a mail invitation with different interest rates on the lender's returns. \cite*{kallus2021fairness} also leverages this dataset to propose policy allocations aimed at maximising the lender's revenue while considering fairness implications. This paper does not explicitly take a stance on fairness. Consequently, we include variables commonly regarded as sensitive in the literature.

The dataset includes 48,852 individuals who received an invitation letter to apply for a loan at the offered interest rate. Differently from the study of \cite{karlan2008credit}, our specification incorporates socio-demographic information (see Table~\ref{varlist_ce} in the Appendix for the complete list) collected from the bank operator before the randomisation. While not crucial for treatment effects identification, these variables are important for exploring heterogeneity and conducting policy learning analysis. In addition, we discretise the treatment variable in three different interest rate groups: low, medium and high rates, based on the client's risk profile. The cut-offs are indicated in \cite{karlan2008credit} and are based on clients' risk category pre-approved by the lender. The discretisation into three groups enhances the interpretability of results from a policy perspective, particularly when employing Policy Trees.

As for the other applications, our heterogeneity analysis is based on the distribution of the IATEs. Specifically, we compare the medium and high-interest rates treatment groups with respect to the low-rate group. As illustrated in Figure \ref{table:policy_tree_ce}, both distributions show positive and negative treatment effects. Looking in more detail at how the medium (high) interest rate group compares to the low-rate group, the effects range from South African Rand R1,520 (R1,656) to R1,610 (R2,327), with an average effect of R78 (R163) and an average standard error of R506 (R487).  These findings indicate the existence of effect heterogeneity, which could be used to potentially enhance the allocation mechanism. 
\input{CE/iates/iates_dens_ce}

As for the previous studies, we use Policy Trees of different depths and compare the welfare produced by the different allocations with the welfare from random and observed allocations.  Due to this application's relatively larger sample size, we also demonstrate the performance of both shallow and deeper trees. Additionally, we explore various combinations of trees: optimal, optimal with one final sequential optimal tree (+1), and optimal with two final sequential optimal trees (+2).

\input{CE/policy_out_ce}

Table~\ref{table:policy_tree_ce} indicates that the observed and random allocations lead to the same welfare since the observed allocations are based on a stratified random assignment. The Table also shows that the Policy Trees increase welfare compared to the baseline policies, irrespective of the tree depth. Even implementing a Policy Tree of depth two demonstrates an increase in overall welfare by 31\%, and deeper trees continue to yield improvements, although at a decreasing rate. The allocation pattern of the optimal Policy Trees shows that deeper trees tend to shift individuals away from the high-rate to a lower-risk group while increasing the average returns at the same time. The welfare estimated with the mixed trees (one or two sequential optimal final trees) is slightly lower than those obtained by the fully optimal tree. For example, the Optimal Policy Tree of depth four generates a welfare of R701, higher than R686 (R685) yielded by the depth-3+1 (depth 2+2) mixed Policy Tree.

Figure~\ref{fig:policy_rules_ce} shows the allocation rule for an optimal Policy Tree of depth 4, which combines both a large welfare improvement and policy rules that are relatively easy to interpret. As indicated in Table~\ref{table:policy_tree_ce}, the Policy Tree assigns roughly half of the sample to the high-rate treatment. 
Among this subgroup, the highest share (16\% of the overall sample) falls between the ages of 35 and 43 and lacks higher education. Low-income individuals older than 53 also constitute a significant portion of the high-risk cohort (12\% of the whole sample). The highest share of individuals assigned to the medium interest rate (14\% of the sample) range from 28 to 32 years of age. The largest subgroup assigned to the low-rate group consists of high-income individuals older than 50 years (4.6\% of the sample).

For detailed information on the treatment allocation rules for the other Policy Trees, we direct the reader to Section \ref{Policy rules Study 3} of the Appendix.

\input{CE/tree_rules_ce}

\section{Technical considerations and conclusion} \label{DC}
An optimal tree of arbitrary depth is computationally challenging due to its NP-hard nature. However, specifying a particular depth allows for polynomial-time resolution, making the depth of the tree a critical complexity parameter \citep*{Sverdrup2020}. Typically, estimating optimal trees beyond a depth of 4 (16 leaves) becomes challenging with moderate-sized training data. To improve computational performance, one option is to reduce the depth, sacrificing some efficiency for computational speed (and improved interpretability). Another option is to use sequential optimal trees. This means building a shallow initial tree, followed by a second shallow tree built from the sub-final leaves of the first one. Building the trees in sequence is much faster than constructing a single, deeper tree of the same overall depth.
Alternatively, the minimum leaf size is crucial for performance optimisation. Setting leaf sizes too small can be impractical and increase computation times.
Additionally, users can enhance speed by reducing the size of the training data at the cost of amplifying sampling noise. Another approach to enhancing speed is to limit the number of values considered during splitting. For continuous variables, the \textit{mcf} algorithm evaluates by default only 100 equally spaced values, while it assesses 100 random combinations for categorical variables. Although this approximation improves computational efficiency, it may lead to some loss in accuracy, particularly as the data in the leaf nodes diminishes. 

Furthermore, like predictive trees, policy trees can also exhibit instability regarding policy rules or welfare estimation. However, how to conduct inference about the stability of policy trees remains an open question in the literature. \cite{athey2021policy} propose a form of cross-validation to evaluate the accuracy of policy learning procedures, and they assess improvements over a random baseline. Similarly, \cite{zhou2023offline} propose an alternative form of cross-validation and test whether optimal policy rules perform significantly better than sending all units to the same treatment. While the assessment of policy tree stability is not currently tackled in the present paper, future research efforts may concentrate on developing methods to evaluate this aspect effectively. 

Despite this, the \textit{mcf} 's \texttt{OptimalPolicy()} class presents a notable advantage by providing a practical implementation of Policy Trees within multi-action settings, facilitating the creation of explainable assignment rules. Moreover, it offers best-scores and random allocation methods, which can be utilised for comparative welfare estimation. Notably, the integration of this class into the \textit{mcf}  package is advantageous. Indeed, it allows for the combination of causal machine learning estimates produced by the \texttt{ModifiedCausalForest()} as inputs for the policy tree, enhancing decision-making. The \textit{mcf} package is an open-source Python package available on the Python Package Index (PyPI). The current work is based on version 0.5.1.

\bibliography{lib}

\section{Appendix} \label{appendix}
\subsection{Codebooks} \label{codebooks}
\input{OHE/codebook_ohe}
\input{RHC/codebook_rhc}
\input{CE/codebook_ce}

\subsection{Policy Trees rules}

This section shows policy rules for individuals allocated to treatment only. All tables indicate the splitting variables and related values produced by our Policy Trees. The codebooks in Section \ref{codebooks} provide the corresponding variable descriptions.
\subsubsection{Policy rules Oregon Health Insurance Experiment} \label{Policy rules Study 1}
\input{OHE/policy_tree_rules_OHE}

\subsubsection{Policy rules Right Heart Catheterization}\label{Policy rules Study 2}
\input{RHC/policy_tables/tree_allocations_rules_wts}
\subsubsection{Policy rules Interest Rates and Loan Returns}\label{Policy rules Study 3}
\input{CE/policy_tree_rules_ce_wts}

\clearpage






\end{document}

%% file: OHE/iates_dens_ohe.tex
\begin{figure}[H]
\centering
\captionsetup{font=small}
\includegraphics[width=0.5\textwidth]{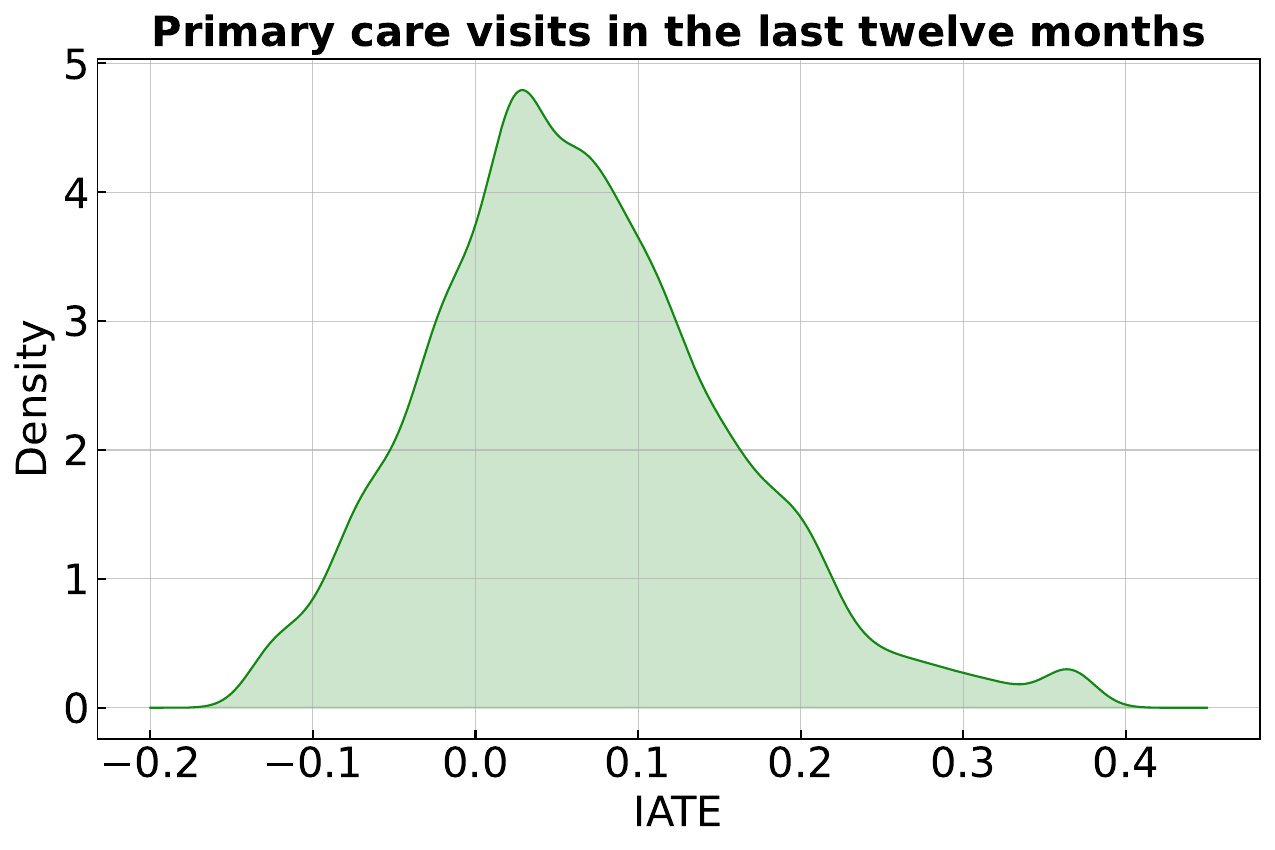}
\caption{Distribution of IATE} 
\label{fig:iates_densities}
\vspace{0.5em}
\centering
\caption*{\textit{Note:} Mean (average standard error) of the IATEs: 0.07 (0.1).}
\end{figure}

%% file: OHE/policy_out_ohe.tex
\begin{table}[ht]
\begin{adjustbox}{width=0.7\columnwidth, center}
\begin{threeparttable}
\captionsetup{font=large}  
    \caption{Treatment allocations Oregon Health Insurance Experiment}
    \centering
        \begin{tabular}{@{}lccc@{}}
    \toprule
    \textbf{Policy} & \textbf{Welfare (\%)} & \multicolumn{2}{c}{\textbf{Treatment shares} (\%)}  \\
            & Primary care visits & Medicaid no & Medicaid yes \\
    \midrule
    Observed & 51.70 & 49.98 & 50.02 \\
    Random & 51.45 & 50.16 & 49.84  \\
    \midrule
    Unconstrained optimal Policy Tree &&& \\
     ~~~~Depth-2 & 54.97 & 0 & 100  \\
     ~~~~Depth-3 & 54.97 & 0  & 100\ \\
    \midrule
     Constrained optimal Policy Tree &&& \\
     ~~~~Depth-2 & 52.88 & 53.56 & 46.44 \\
     ~~~~Depth-3 & 53.48 & 51.75 & 48.25 \\
     ~~~~Depth-4 & 53.51 & 52.18 & 47.82 \\
    \midrule
   Constrained sequentially optimal Policy Trees &&& \\
     ~~~~Depth-2+1 &  52.68 & 57,06 & 42,93 \\
     ~~~~Depth-3+1 &  53.54 & 50.30 & 49.69\\
    \bottomrule
    \end{tabular}
    \begin{tablenotes}
    \textit{Note:} 
    The table shows empirically observed, randomly assigned, and optimal treatment allocations of individuals. The optimal policy trees are unconstrained and constrained. The final constrained trees add a sequential optimal sub-tree (+1). The constraint is implemented as a limit imposed on the treatment shares, as observed within the empirical dataset. The outcome represents primary care visits. The treatment is being extracted from the lottery, which allows one to apply for Medicaid, the public health insurance programme. 
    \end{tablenotes}
    \label{table:policy_tree_ohe}
\end{threeparttable}
\end{adjustbox}
\end{table}

%% file: OHE/policy_tree_ohe.tex
\begin{figure}[H]
\captionsetup{font=small}  
\centering
\includegraphics[width=0.4\textwidth]{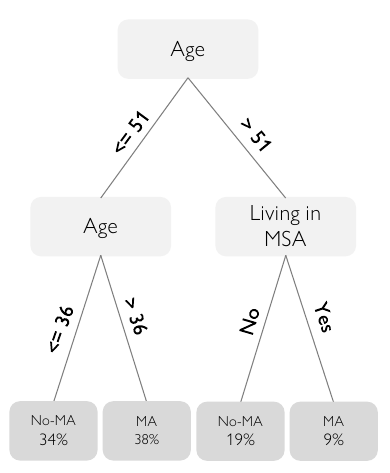}
\caption{Allocation rule for a constrained optimal Policy Tree of depth-2}
\caption*{\textit{Note:} The policy tree shows the shares of individuals allocated to each treatment. The splitting variables of the tree are age and an indicator variable for living in a metropolitan area (MSA). The final leaves of the trees indicate the treatments and corresponding shares of allocated individuals. The treatment used in the analysis is the indicator for being assigned to the lottery, which consequently gives access to Medicaid. For representation reasons, the tree reports: No-MA: No access to Medicaid (not extracted for the lottery), MA: Access to Medicaid (extracted for the lottery).}
\label{fig:policy_rules_ohe}
\end{figure}

%% file: RHC/IATE/iates_density.tex
\begin{figure}[H]
\centering
\captionsetup{font=small}  
\includegraphics[width=0.5\textwidth]{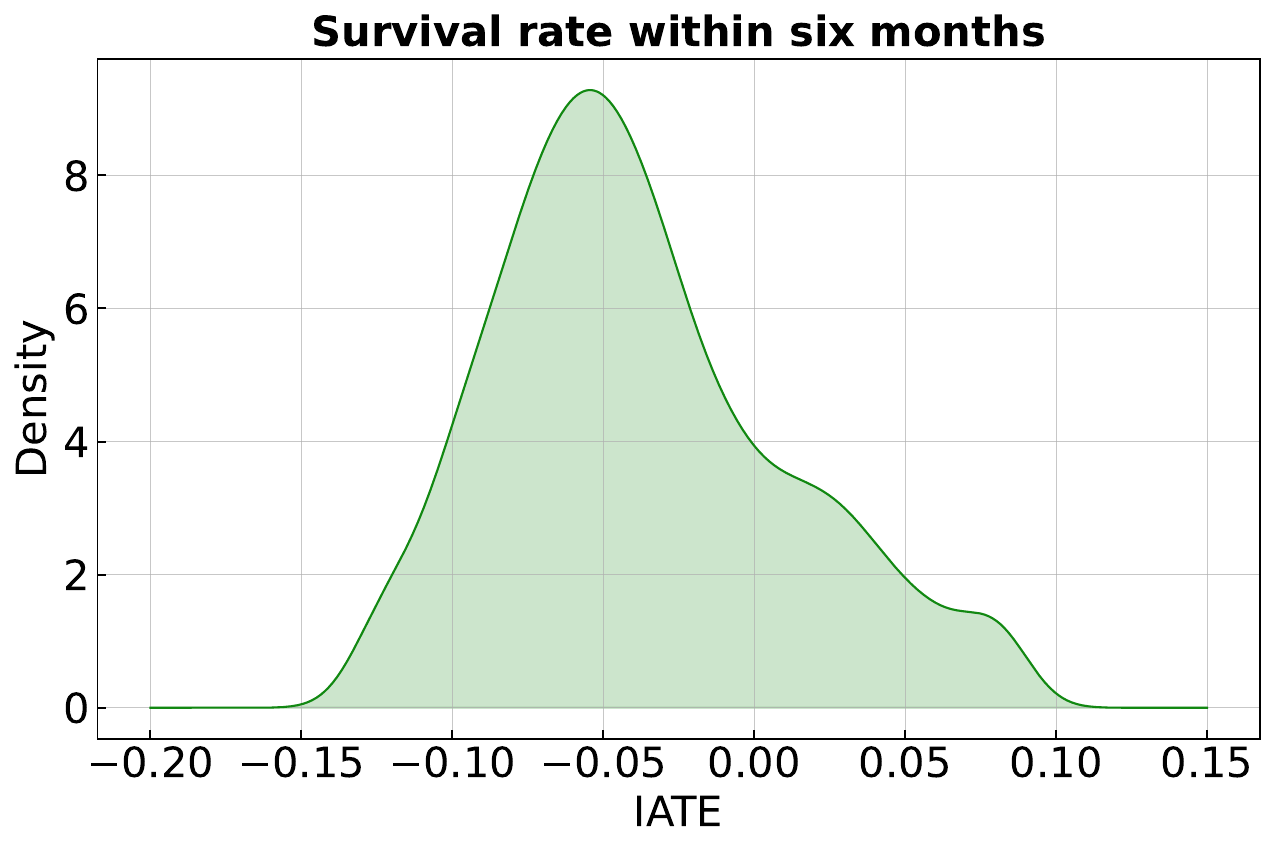}
\caption{Distribution of IATE}
\caption*{\textit{Note:} Mean (average standard error) of the IATEs: -0.04 (0.06).}
\label{fig:iates_dens_rhc}
\end{figure}


%% file: RHC/policy_tables/policy_out.tex
\begin{table}[ht]
\begin{adjustbox}{width=0.7\columnwidth, center}
\begin{threeparttable}
\captionsetup{font=large}  
    \caption{Treatment allocations  Right Heart Catheterization}
    \centering
    \begin{tabular}{@{}lrrr@{}}
    \toprule
    \textbf{Policy} & \textbf{Welfare (\%)} & \multicolumn{2}{c}{\textbf{Treatment shares (\%)}}  \\
        & Survival rate & no RHC & yes RHC \\

    \midrule
    Observed & 68.23 & 60.51 & 39.49 \\ 
    Random &  67.88  & 59.00 & 40.99 \\ 
    \midrule
    Unconstrained optimal Policy Tree  &&& \\
    ~~~~Depth-2 & 69.99 & 80.78 & 19.21\\
    ~~~~Depth-3 & 69.95 & 82.73 & 17.27 \\
    \midrule
   Unconstrained sequentially optimal Policy Trees &&& \\
    ~~~~Depth-2+1 &  69.98 & 82.88 & 17.11 \\
    ~~~~Depth-3+1 &  69.97 & 83.78  &16.22 \\
    \bottomrule
    \end{tabular}
    \begin{tablenotes}
    \textit{Note:} 
    The table displays empirically observed, randomly assigned, and optimised treatment allocations as policies, where the final two Policy Trees add a second optimal sub-tree (+1).
    \end{tablenotes}
    \label{table:policy_tree_rhc}
\end{threeparttable}
\medskip
\end{adjustbox}
\end{table}

%% file: RHC/policy_tables/rhc_tree.tex
\begin{figure}[H]
\centering
\captionsetup{font=small}  
\includegraphics[width=0.4\textwidth]{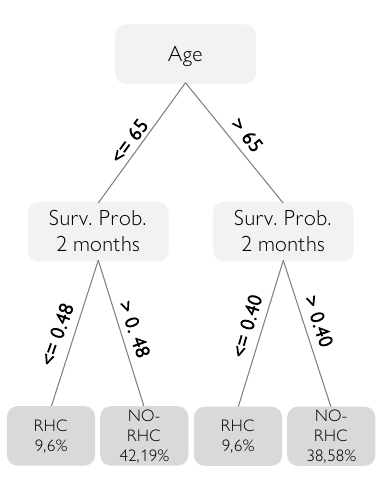}
\caption{Allocation rule for an optimal Policy Tree of depth-2}
\caption*{\textit{Note:} The policy tree shows the shares of individuals allocated to treatment. The split variables are age and the probability of surviving within two months. The variable is measured before the decision of surgery. The final leaves of the trees indicate the treatments and corresponding shares of individuals allocated.}
\label{fig:tree_alloc_rules}
\end{figure}

%% file: CE/iates/iates_dens_ce.tex
\begin{figure}[H]
\centering
\captionsetup{font=small}  
\includegraphics[width=0.5\textwidth]{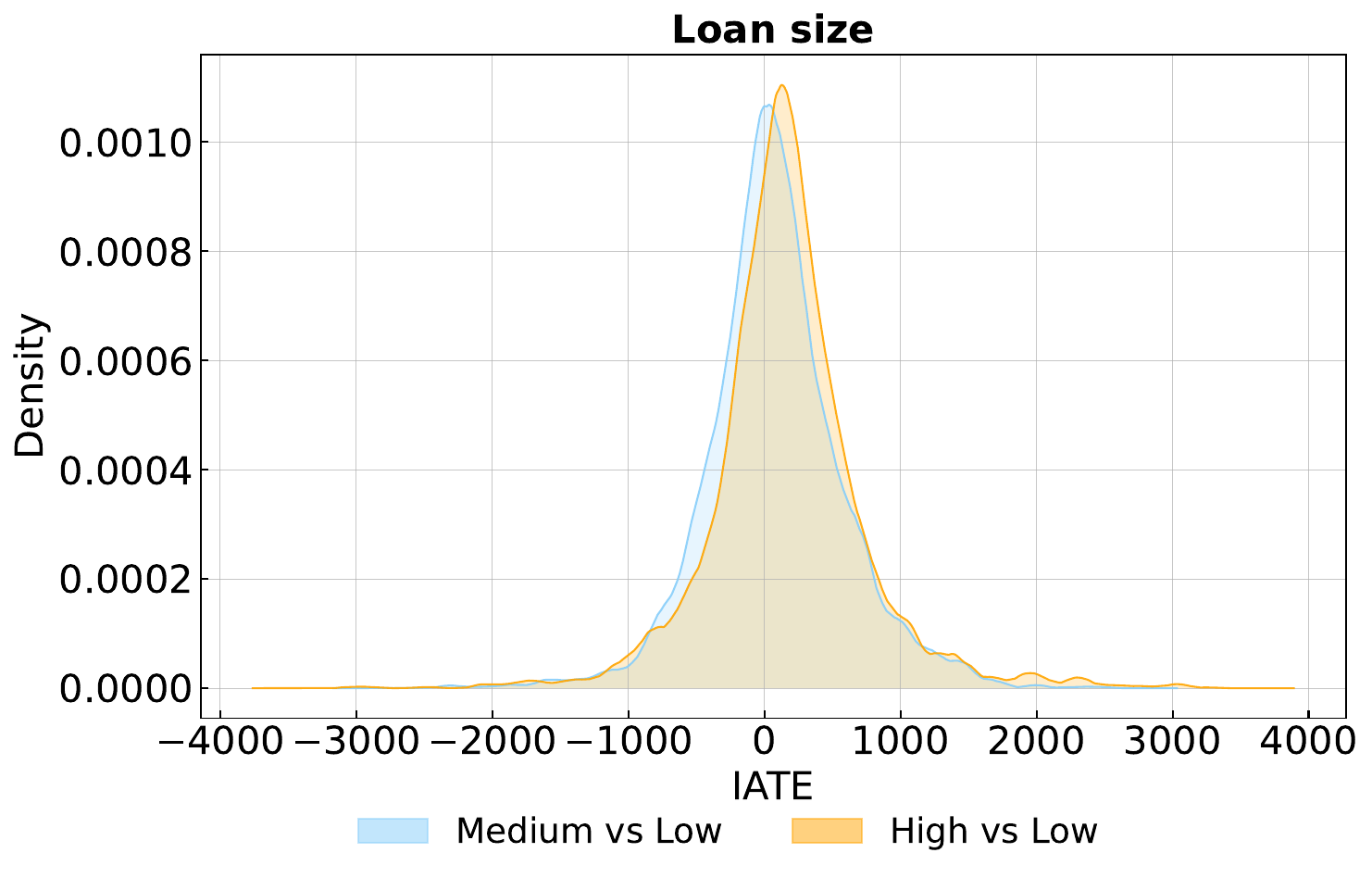}
\caption{Distribution of IATE}
\caption*{\textit{Note:} Means of the IATEs for the high (medium) vs low-interest rate groups: R78 (R163). Average standard errors of the IATEs for the high (medium) vs low-interest rate groups: R506 (R487).}
\label{fig:iates_dens_ce}
\end{figure}

%% file: CE/policy_out_ce.tex
\begin{table}[ht]
\begin{adjustbox}{width=0.7\columnwidth, center}
\begin{threeparttable}
\captionsetup{font=large}  
    \caption{Treatment allocations Interest Rates and Loan Returns}
    \centering
        \begin{tabular}{@{}lcccc@{}}
    \toprule
    \textbf{Policy} & \textbf{Welfare (R)} & \multicolumn{3}{c}{\textbf{Treatment shares (\%)}}  \\
    & Average returns & Low rate & Medium rate & High rate \\
    \midrule
    Observed & 497 & 39.84 & 28.40 & 31.75 \\ 
    Random &  497 & 39.84  & 28.85 & 31.10\\ 
    \midrule
     Unconstrained optimal Policy Tree && & &\\
     ~~~~Depth-2 & 652 & 4.62 & 40.25 & 55.12\\ 
     ~~~~Depth-3 & 678 & 4.62 & 44.62 & 50.76 \\
     ~~~~Depth-4 & 701 & 13.74 & 36.02 & 50.23 \\
    \midrule
    Unconstrained sequentially optimal Policy Trees && & &\\
     ~~~~Depth-2+1 & 663 & 19.58 & 28.89 & 51.53 \\
     ~~~~Depth-3+1 & 686 &  20.95 & 25.14 & 53.90 \\
     ~~~~Depth-4+1 & 703 & 11.68 & 31.23 & 57.09 \\
     ~~~~Depth-2+2 & 685 & 20.84 & 28.67 & 50.50 \\
     ~~~~Depth-3+2 & 710 & 14.74 & 34.66 & 50.60 \\
     ~~~~Depth-4+2 & 717 & 17.45 & 28.87 & 53.68 \\
    \bottomrule
    \end{tabular}
    \begin{tablenotes}
    \textit{Note:} 
    The table shows the simulated and observed allocations of individuals per program under different policies. The outcome represents the average loan size in South African Rands (R7 = 1 USD).
    \end{tablenotes}
    \label{table:policy_tree_ce}
\end{threeparttable}
\end{adjustbox}
\end{table}

%% file: CE/tree_rules_ce.tex
\begin{figure}[H]
\captionsetup{font=small}  
\centering
\includegraphics[width=1\textwidth]{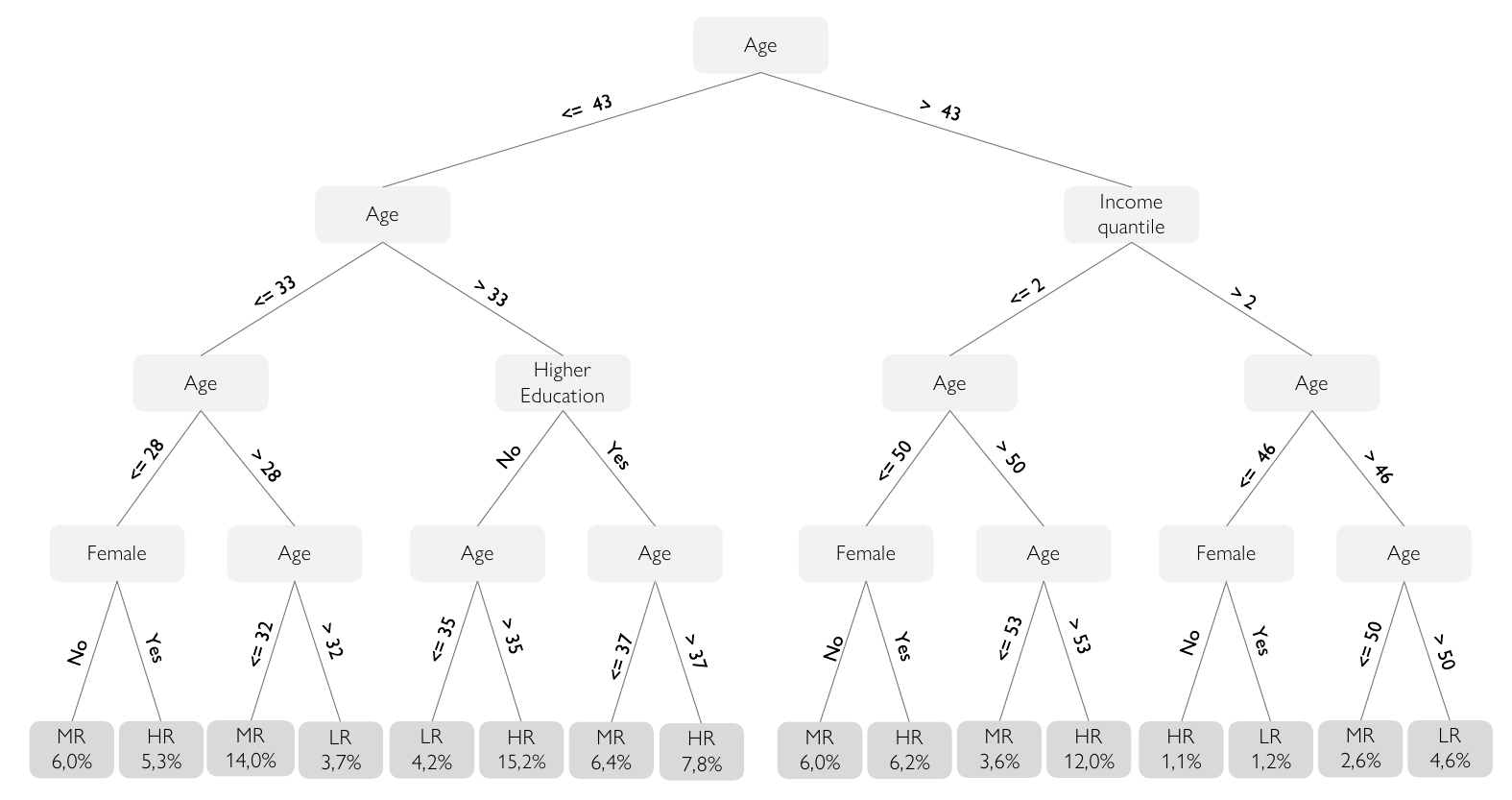} 
\caption{Allocation rule for an optimal Policy Tree of depth-4}
\caption*{\textit{Note:} The policy tree shows the shares of individuals allocated to each treatment. The splitting variables of the tree are gender, age, income quantile and an indicator for achieved higher education. The final leaves of the trees indicate the treatments and corresponding shares of allocated individuals. LR: Low rate, MR: Medium rate, HR: High rate.}
\label{fig:policy_rules_ce}
\end{figure}

%% file: OHE/codebook_ohe.tex
\begin{xltabular}{\textwidth}{lX}
\caption{Codebook for the Oregon health insurance study (Section \ref{OHE})} \\
\toprule \multicolumn{1}{l}{\textbf{Variable}} & \multicolumn{1}{l}{\textbf{Description}} \\ \midrule 
\endfirsthead

\multicolumn{2}{c}%
{\tablename\ \thetable{} -- continued from previous page} \\
\hline \multicolumn{1}{l}{\textbf{variable}} & \multicolumn{1}{l}{\textbf{description}} \\ \hline 
\endhead

\hline \multicolumn{2}{r}{{continued on next page}} \\ \hline
\endfoot

\bottomrule
\endlastfoot
\label{varlist_ohe}

doc\_any\_12m & any primary care visits: survey data 12m \\
treatment & indicator equal to 1 if extracted for the lottery \\
age & in years, calculated as difference between 2008 and birth year \\
numhh\_list & number of people in household on lottery list \\
draw & survey mailing wave for 0m survey \\
birthyear\_list & birth year: lottery list data \\
have\_phone\_list & gave a phone number on lottery sign up: lottery list data \\
pobox\_list & gave a PO Box as an address: lottery list data \\
english\_list & individual requested english-language materials: lottery list data \\
female\_list & indicator for female: lottery list data \\
zip\_msa & zip code from lottery list is a metropolitan statistical area\\

\end{xltabular}

%% file: RHC/codebook_rhc.tex
\begin{xltabular}{\textwidth}{lX}
\caption{Codebook for the Right Heart Catheterization (Section \ref{RHC})} \\
\toprule \multicolumn{1}{l}{\textbf{Variable}} & \multicolumn{1}{l}{\textbf{Description}} \\ \midrule 
\endfirsthead

\multicolumn{2}{c}%
{\tablename\ \thetable{} -- continued from previous page} \\
\hline \multicolumn{1}{l}{\textbf{Variable}} & \multicolumn{1}{l}{\textbf{Description}} \\ \hline 
\endhead

\hline \multicolumn{2}{r}{{continued on next page}} \\ \hline
\endfoot

\bottomrule
\endlastfoot
\label{varlist_rhc}
    adld3pc & index of activities of daily living 2 weeks prior to admission \\
    age   & age in years \\
    alb1  & albumin level in grams per deciliter \\
    amihx & indicator for definite myocardial infarction \\
    aps1  & acute physiology and chronic health evaluation score (APACHE III) \\
    bili1 & bilirubin level in milligrams per deciliter \\
    ca & cancer; 3 categories: metastatic cancer (Metastatic), no cancer (No), cancer (Yes) \\
    card  & indicator for cardiovascular diagnosis \\
    cardiohx & indicator for acute myocardial infarction, peripheral vascular disease, severe, and very severe cardiovascular symptoms (NYHA-classes III and IV) \\
    cat1 & primary disease; 9 categories: acute respiratory failure (0), congestive heart failure (1), chronic obstructive pulmonary disease (2), cirrhosis (3), colon cancer metastatic to the liver (4), non-traumatic coma (5), non-small-cell cancer of the lung, stage III or IV (6), multiorgan system failure with malignancy (7), indicator for multiorgan system failure with sepsis (8) \\
    cat2 & secondary disease; 3 categories: missings or others (0), multiorgan system failure with malignancy (1), multiorgan system failure with sepsis(2) \\
	cat2\_miss & indicator for missings in cat2 \\
    chfhx & indicator for congestive heart failure \\
    chrpulhx & indicator for chronic, severe, or very severe pulmonary disease \\
    crea1 & creatinine level in milligrams per deciliter \\
    das2d3pc & Duke activity status index (DASI) two weeks before admission \\
    dementhx & indicator for dementia, stroke or cerebral infarct, Parkinson’s disease \\
    dnr1  & indicator for do not resuscitate status on the first day \\
    dth30 & outcome indicator for mortality within 6 months \\
    edu   & education in years \\
    gastr & indicator for gastrointestinal diagnosis \\
    gibledhx & indicator for upper gastrointestinal bleeding \\
    hema  & indicator for hematologic diagnosis \\
    hema1 & hematocrit levels in percent \\
    hrt1  & heart rate in beats per minute \\
    immunhx & indicator for immunosuppression, organ transplant, HIV positivity, diabetes mellitus with or without end organ damage, connective tissue disease \\
    income & income in US dollars; 4 categories:  below 11k (0), 11-25k (1), 25-50k (2), above 50k (3) \\
    liverhx & indicator for cirrhosis, hepatic failure \\
    malighx & indicator for solid tumor, metastatic disease, chronic leukemia/myeloma, acute leukemia, lymphoma \\
    meanbp1 & mean blood pressure in millimeters of mercury \\
    meta  & indicator for metabolic diagnosis \\
    neuro & indicator for neurological diagnosis \\
    ninsclas & medical insurance; 6 categories: Medicaid (0), Medicare (1), Medicare and Medicaid (2), no medical insurance (3), private medical insurance (4), private medical insurance and Medicaid (5)\\
    ortho & indicator for orthopedic diagnosis \\
    paco21 & partial pressure of arterial carbon dioxide in millimeters of mercury \\
    pafi1 & ratio between partial pressure of arterial oxygen in millimeters of mercury and fraction of inspired oxygen \\
    ph1   & potential of hydrogen (pH) at logarithmic scale \\
    pot1  & blood potassium level in millimoles per liter \\
    psychhx & indicator for psychiatric history, active psychosis or severe depression \\
    race  & race; 3 categories: African American (0), other race (1), Caucasian (2) \\
    renal & indicator for renal diagnosis \\
    renalhx & indicator for chronic renal disease, chronic hemodialysis, or peritoneal dialysis \\
    resp  & indicator for respiratory diagnosis \\
    resp1 & respiratory rate \\
    scoma1 & Glasgow coma score \\
    seps  & indicator for sepsis diagnosis \\
    sex   & indicator for gender (1=male) \\
    sod1  & sodium level in milliequivalents per liter \\
    surv2md1 & probability of surviving two months (based on support model estimations) \\
    swang1 & treatment indicator for right heart catheterisation \\
    temp1 & body temperature in degrees Celsius \\
    transhx & indicator for transfer exceeding 24 hours from another hospital \\
    trauma & indicator for trauma diagnosis \\
    urin1 & urine output per day in milliliters \\
		urin1\_miss & indicator for missings in urin1 \\
    wblc1 & white blood cells in thousands per cubic millimeter \\
    wtkilo1 & weight in kilograms \\

\end{xltabular}

%% file: CE/codebook_ce.tex
\begin{xltabular}{\textwidth}{lX}
\caption{Codebook for the Interest Rates and Loan Returns study (Section \ref{CE})} \\
\toprule \multicolumn{1}{l}{\textbf{Variable}} & \multicolumn{1}{l}{\textbf{Description}} \\ \midrule 
\endfirsthead

\multicolumn{2}{c}%
{\tablename\ \thetable{} -- continued from previous page} \\
\hline \multicolumn{1}{l}{\textbf{Variable}} & \multicolumn{1}{l}{\textbf{Description}} \\ \hline 
\endhead

\hline \multicolumn{2}{r}{{continued on next page}} \\ \hline
\endfoot

\bottomrule
\endlastfoot
\label{varlist_ce}

adj\_loan & loan size (loansize variable) multiplied by the interest rate (final4 variable) applied to calculate returns.\\
offer\_risk & treatment, based on the original offered interest rate (offer4); offer\_risk is equal to 1 for the offers 
              with a value less than or equal to 7.75, indicating an associated low risk; offers with values greater than 
              7.75 but less than or equal to 9.75 are categorized as 2, associated to moderate risk; offers with a value 
              greater than 9.75 are categorized as 3, associated to high risk. \\
age & age in years \\
dependants & number of individuals who rely on the primary household for economic support and care \\
qnt\_income & quintiles of gross income \\
wave & number indicating in which wave the offer letter has been sent\\
risk\_num & number from 1 to 3 indicating the client's risk category\\
female & indicator for gender, equal to 1 if female \\
married & indicator for civil status, equal to 1 if married\\
edhi & indicator for higher education, equal to 1 if this level is achieved\\
rural & indicator for living in rural area\\

\end{xltabular}

%% file: OHE/policy_tree_rules_OHE.tex
\begin{table}[H]
\centering
\caption{Unconstrained optimal Policy Tree of depth-2}
\begin{flushleft}
\begin{tabular}{l}
\toprule
Splitting variables and values \\ \hline
female\_list = 0, numhh\_list = 1  \\ \hline
female\_list = 0, numhh\_list $>$ 1  \\ \hline
female\_list = 0, age $\leq$ 47 \\ \hline
female\_list = 0, age $>$ 47 \\ 
\bottomrule
\end{tabular}
\end{flushleft}
\begin{tablenotes}
\footnotesize 
\textit{Note:} Policy rule for individuals assigned to Medicaid.
\end{tablenotes}
\end{table}

\begin{table}[H]
\centering
\caption{Unconstrained optimal Policy Tree of depth-3}
\begin{flushleft}
\begin{tabular}{l}
\toprule
Splitting variables and values \\ \hline
age $\leq$ 50, age $\leq$ 31, age $\leq$ 26 \\ \hline
age $\leq$ 50, age $\leq$ 31, age $>$ 26  \\ \hline
age $\leq$ 50, age $>$ 31, age $\leq$ 38\\ \hline
age $\leq$ 50, age $>$ 31, age $>$ 38  \\ \hline
age $>$ 50, numhh\_list = 1, age $\leq$ 60  \\ \hline
age $>$ 50, numhh\_list = 1, age $>$ 60  \\ \hline
age $>$ 50, numhh\_list $>$ 1, age $\leq$ 59  \\ \hline
age $>$ 50, numhh\_list $>$ 1, age $>$ 59 \\ 
\bottomrule
\end{tabular}
\end{flushleft}
\begin{tablenotes}
\footnotesize 
\textit{Note:} Policy rule for individuals assigned to Medicaid.
\end{tablenotes}
\end{table}

\begin{table}[H]
\caption{Constrained optimal Policy Tree of depth-2}
\begin{flushleft}
\begin{tabular}{l}
\toprule
Splitting variables and values \\
\midrule
age $\leq$ 51, age $>$ 36 \\ \hline
age $>$ 51, numhh\_zip\_msa = 0 \\
\bottomrule
\end{tabular}
\end{flushleft}
\begin{tablenotes}
\footnotesize 
\textit{Note:} Policy rule for individuals assigned to Medicaid.
\end{tablenotes}
\end{table}

\begin{table}[H]
\caption{Unconstrained Optimal Policy Tree of depth-3}
\begin{flushleft}
\begin{tabular}{l}
\toprule
Splitting variables and values \\
\midrule
numhh\_list = 1, age $\leq$ 51, age $>$ 37 \\ \hline
numhh\_list = 1, age $>$ 51, zip\_msa = 0  \\ \hline
numhh\_list $>$ 1, female\_list = 0, age $>$ 25 \\ \hline
numhh\_list $>$ 1, female\_list = 1, zip\_msa = 0\\ 
\bottomrule
\end{tabular}
\end{flushleft}
\begin{tablenotes}
\footnotesize 
\textit{Note:} Policy rule for individuals assigned to Medicaid.
\end{tablenotes}
\end{table}

\begin{table}[H]
\caption{Constrained sequentially optimal Policy Trees of depth-2+1}
\begin{flushleft}
\begin{tabular}{l}
\toprule
Splitting variables and values \\ 
\midrule
age $\leq$ 50, age $>$ 39, age $\leq$ 40 \\ \hline
age $\leq$ 50, age $>$ 39, age $>$ 40  \\ \hline
age $>$ 50, numhh\_zip\_msa = 0, age $\leq$ 53 \\ \hline
age $>$ 50, numhh\_zip\_msa = 0, age $>$ 53 \\ \hline
age $>$ 50, zip\_msa = 1 numhh\_list $>$ 1\\ 
\bottomrule
\end{tabular}
\end{flushleft}
\begin{tablenotes}
\footnotesize 
\textit{Note:} Policy rule for individuals assigned to Medicaid.
\end{tablenotes}
\end{table}

\begin{table}[H]
\centering
\caption{Constrained optimal Policy Tree of depth-4}
\begin{flushleft}
\begin{tabular}{l}
\toprule
numhh\_list = 1, zip\_msa = 0, female\_list = 0, age $>$ 28 \\ \hline
numhh\_list = 1, zip\_msa = 0, female\_list $>$ 0.500, age $>$ 35  \\ \hline
numhh\_list = 1, zip\_msa = 1, age $>$ 39, age $\leq$ 51 \\ \hline
numhh\_list $>$ 1, female\_list = 0, zip\_msa < 0, age $\leq$ 36 \\ \hline
numhh\_list $>$ 1, female\_list = 0 zip\_msa = 0, age $>$ 36\\ \hline
numhh\_list $>$ 1, female\_list = 0, zip\_msa = 1, age $>$ 28 \\ \hline
numhh\_list $>$ 1, female\_list = 1, english\_list = 0, age $\leq$ 42 \\ \hline
numhh\_list $>$ 1, female\_list = 1, english\_list = 1, zip\_msa = 0 \\ \hline

\end{tabular}
\end{flushleft}
\begin{tablenotes}
\footnotesize 
\textit{Note:} Policy rule for individuals assigned to Medicaid.
\end{tablenotes}
\end{table}

\begin{table}[H]
\centering
\caption{Constrained sequentially optimal Policy Tree of depth-3+1}
\begin{flushleft}
\begin{tabular}{l}
\toprule
Splitting variables and values \\
\midrule
numhh\_list $>$ 1, female\_list = 0, age $>$ 27 \\ \hline
numhh\_list = 1, age $\leq$ 51, age $>$ 37, age $<$ 39 \\ \hline
numhh\_list = 1, age $\leq$ 51, age $>$ 37, age $>$ 39 \\ \hline
numhh\_list = 1, age $>$ 51, zip\_msa = 0, age $\leq$ 53  \\ \hline
numhh\_list = 1, age $>$ 51, zip\_msa = 0, age $>$ 53 \\ \hline
numhh\_list $>$ 1, female\_list = 1, zip\_msa = 0, age $\leq$ 53 \\ \hline
numhh\_list $>$ 1, female\_list = 1, zip\_msa = 0, age $>$ 53\\ \hline
numhh\_list $>$ 1, female\_list = 1, zip\_msa = 1, english\_list = 0  \\ \hline
\end{tabular}
\end{flushleft}
\begin{tablenotes}
\footnotesize 
\textit{Note:} Policy rule for individuals assigned to Medicaid.
\end{tablenotes}
\end{table}

%% file: RHC/policy_tables/tree_allocations_rules_wts.tex
\begin{table}[H]
\caption{Unconstrained optimal Policy Tree of depth-2}
\begin{flushleft}    
\begin{tabular}{l}
\toprule
Splitting variables and values   \\ \midrule
age $\leq$ 65, surv2md1 $\leq$ 0.480 \\
age $>$ 65, surv2md1 $\leq$ 0.402  \\ \bottomrule
\end{tabular}
\end{flushleft}
\begin{tablenotes}
\footnotesize
\textit{Note:} Policy rule for individuals assigned to RHC.
\end{tablenotes}
\end{table}

\begin{table}[H]
\caption{Unconstrained optimal Policy Tree of depth-3}
\begin{flushleft}
\begin{tabular}{l}
\toprule
Splitting variables and values \\ \midrule
cat1 in: 0 1 2 6 8, age $\leq$ 70, surv2md1 $\leq$ 0.450 \\ \hline
cat1 in: 0 1 2 6 8, age $>$ 70, surv2md1 $\leq$ 0.402  \\ \hline
cat1 not in: 0 1 2 6 8, cat1 in: 3 5, aps1 $>$ 64\\ \hline
cat1 not in: 0 1 2 6 8, cat1 not in: 3 5, surv2md1 $\leq$ 0.467  \\ \bottomrule
\end{tabular}
\end{flushleft}
\begin{tablenotes}
\footnotesize
\textit{Note:} Policy rule for individuals assigned to RHC.
\end{tablenotes}
\end{table}

\begin{table}[H]
\caption{Unconstrained sequentially optimal Policy Trees of depth-2+1}
\begin{flushleft}
\begin{tabular}{l}
\toprule
Splitting variables and values   \\ \midrule
age $\leq$ 65, surv2md1 $\leq$ 0.480, aps1 $\leq$ 80 \\ \hline
age $\leq$ 65, surv2md1 $\leq$ 0.480, aps1 $>$ 80 \\ \hline
age $>$ 65, surv2md1 $\leq$ 0.402, aps1 $>$ 50 \\ \bottomrule
\end{tabular}
\end{flushleft}
\begin{tablenotes}
\footnotesize
\textit{Note:} Policy rule for individuals assigned to RHC.
\end{tablenotes}
\end{table}

\begin{table}[H]
\caption{Unconstrained sequentially optimal Policy Trees of depth-3+1}
\begin{flushleft}
\begin{tabular}{l}
\toprule
Splitting variables and values  \\ \midrule
cat1 in: 0 2 8, age $\leq$ 69, surv2md1 $\leq$ 0.450, cat1 = 0\\ \hline
cat1 in: 0 2 8, age $\leq$ 69, surv2md1 $\leq$ 0.450, cat1 not in: 0  \\ \hline
cat1 in: 0 2 8, age $>$ 69, surv2md1 $\leq$ 0.402, aps1 $>$ 61 \\ \hline
cat1 not in: 0 2 8, cat1 in: 3 5, aps1 $>$ 64, aps1 $\leq$ 77 \\ \hline
cat1 not in: 0 2 8, cat1 in: 3 5, aps1 $>$ 64, aps1 $>$ 77  \\ \hline
cat1 not in: 0 2 8, cat1 not in: 3 5, surv2md1 $\leq$ 0.467, surv2md1 $\leq$ 0.384 \\ \hline
cat1 not in: 0 2 8, cat1 not in: 3 5, surv2md1 $\leq$ 0.467, surv2md1 $>$ 0.384 \\ \bottomrule
\end{tabular}
\end{flushleft}
\begin{tablenotes}
\footnotesize
\textit{Note:} Policy rule for individuals assigned to RHC.
\end{tablenotes}
\end{table}

%% file: CE/policy_tree_rules_ce_wts.tex
\begin{table}[ht]
\caption{Unconstrained Optimal Policy Tree of depth-2}
\begin{flushleft}
\begin{tabular}{l|c}
\toprule
Splitting variables and values & Treatment allocation \\ \midrule
age $\leq$ 50, age $>$ 36 & HR  \\ \hline
age $>$ 50, qnt\_income $\leq$ 2 & HR  \\ \hline
age $\leq$ 50, age $\leq$ 36 & MR  \\ \hline
age $>$ 50, qnt\_income $>$ 2 & LR \\ \bottomrule
\end{tabular}
\end{flushleft}
\begin{tablenotes}
\footnotesize
\textit{Note:} Policy rule for individuals assigned to HR: high rate, MR: medium rate and LR: low rate.
\end{tablenotes}
\end{table}

\begin{table}[ht]
\caption{Unconstrained optimal Policy Tree of depth-3}
\begin{flushleft}
\begin{tabular}{l|c}
\bottomrule
Splitting variables and values & Treatment allocation \\ \midrule
age $>$ 46, qnt\_income $>$ 2, age $>$ 50 & LR \\ \hline
age $\leq$ 46, age $\leq$ 28, female = 0 & MR  \\ \hline
age $\leq$ 46, age $\leq$ 28, female = 1 & HR  \\ \hline
age $\leq$ 46, age $>$ 28, age $\leq$ 36 & MR \\ \hline
age $>$ 46, qnt\_income $\leq$ 2, age $\leq$ 53 & MR\\ \hline
age $\leq$ 46, age $>$ 28, age $>$ 36 & HR  \\ \hline
age $>$ 46, qnt\_income $\leq$ 2, age $>$ 53 & HR \\ \hline
age $>$ 46, qnt\_income $>$ 2, age $\leq$ 50 & HR \\ \bottomrule
\end{tabular}
\end{flushleft}
\begin{tablenotes}
\footnotesize
\textit{Note:} Policy rule for individuals assigned to HR: high rate, MR: medium rate and LR: low rate.
\end{tablenotes}
\end{table}

\begin{table}[ht]
\caption{Unconstrained optimal Policy Tree of depth-4}
\begin{flushleft}
\begin{tabular}{l|c}
\toprule
Splitting variables and values & Treatment allocation \\ \midrule
age $>$ 43, qnt\_income $>$ 2, age $>$ 46, age $>$ 50 & LR  \\ \hline
age $\leq$ 43, age $\leq$ 33, age $>$ 28, age $>$ 32 & LR  \\ \hline
age $\leq$ 43, age $>$ 33, edhi = 0, age $\leq$ 35& LR  \\ \hline
age $>$ 43, qnt\_income $>$ 2, age $\leq$ 46, female =1& LR \\ \hline
age $\leq$ 43, age $\leq$ 33, age $\leq$ 28 female =1& MR  \\ \hline
age $\leq$ 43, age $\leq$ 33, age $>$ 28, age $\leq$ 32 & MR  \\ \hline
age $\leq$ 43, age $>$ 33, edhi =1, age $\leq$ 37& MR \\ \hline
age $>$ 43, qnt\_income $\leq$ 2, age $\leq$ 50, female = 0& MR  \\ \hline
age $>$ 43, qnt\_income $\leq$ 2, age $>$ 50, age $\leq$ 53 & MR  \\ \hline
age $\leq$ 43, age $\leq$ 33, age $\leq$ 28, female = 0& HR \\ \hline
age $\leq$ 43, age $>$ 33, edhi = 0, age $>$ 35& HR  \\ \hline
age $\leq$ 43, age $>$ 33, edhi = 1, age $>$ 37& HR \\ \hline
age $>$ 43, qnt\_income $\leq$ 2, age $\leq$ 50, female = 1& HR \\ \hline
age $>$ 43, qnt\_income $\leq$ 2, age $>$ 50, age $>$ 53 & HR  \\ \hline
age $>$ 43, qnt\_income $>$ 2, age $\leq$ 46, female = 0& HR \\ \hline
age $>$ 43, qnt\_income $>$ 2, age $>$ 46, age $\leq$ 50 & HR  \\ \bottomrule
\end{tabular}
\end{flushleft}
\begin{tablenotes}
\footnotesize
\textit{Note:} Policy rule for individuals assigned to HR: high rate, MR: medium rate and LR: low rate.
\end{tablenotes}
\end{table}

\begin{table}[ht]
\caption{Unconstrained sequentially optimal Policy Trees of depth-2+1}
\begin{flushleft}    
\begin{tabular}{l|c}
\toprule
Splitting variables and values & Treatment allocation \\ \midrule
age $\leq$ 50, age $\leq$ 36, age $>$ 32 & LR \\ \hline
age $>$ 50, qnt\_income $>$ 2, age $\leq$ 57 & LR \\ \hline
age $>$ 50, qnt\_income $>$ 2, age $>$ 57 & LR \\ \hline
age $\leq$ 50, age $\leq$ 36, age $\leq$ 32 & MR  \\ \hline
age $>$ 50, qnt\_income $\leq$ 2, age $\leq$ 53 & MR \\ \hline
age $\leq$ 50, age $>$ 36, age $\leq$ 38 & HR \\ \hline
age $\leq$ 50, age $>$ 36, age $>$ 38 & HR \\ \hline
age $>$ 50, qnt\_income $\leq$ 2, age $>$ 53 & HR\\ \bottomrule
\end{tabular}
\end{flushleft}
\begin{tablenotes}
\footnotesize
\textit{Note:} Policy rule for individuals assigned to HR: high rate, MR: medium rate and LR: low rate.
\end{tablenotes}
\end{table}

\begin{table}[ht]
\caption{Unconstrained sequentially optimal Policy Trees of depth-3+1}
\begin{flushleft}    
\begin{tabular}{l|c}
\toprule
Splitting variables and values & Treatment allocation \\
\midrule
age $\leq$ 46, age $\leq$ 28, female = 0, age $\leq$ 27& MR   \\
\hline
age $\leq$ 46, age $\leq$ 28, female = 0., age $>$ 27& HR  \\
\hline
age $\leq$ 46, age $\leq$ 28, female = 1, age $\leq$ 25& HR \\
\hline
age $\leq$ 46, age $\leq$ 28, female = 1, age $>$ 25& HR  \\
\hline
age $\leq$ 46, age $>$ 28, age $\leq$ 36, age $\leq$ 32 & MR  \\
\hline
age $\leq$ 46, age $>$ 28, age $\leq$ 36, age $>$ 32 & LR \\
\hline
age $\leq$ 46, age $>$ 28, age $>$ 36, age $\leq$ 37 & HR \\
\hline
age $\leq$ 46, age $>$ 28, age $>$ 36, age $>$ 37 & HR  \\
\hline
age $>$ 46, qnt\_income $\leq$ 2, age $\leq$ 53, qnt\_income = 0& HR \\
\hline
age $>$ 46, qnt\_income $\leq$ 2, age $\leq$ 53, qnt\_income $>$ 0 & MR  \\
\hline
age $>$ 46, qnt\_income $\leq$ 2, age $>$ 53, rural = 0& HR  \\
\hline
age $>$ 46, qnt\_income $\leq$ 2, age $>$ 53, rural = 1& HR  \\
\hline
age $>$ 46, qnt\_income $>$ 2, age $\leq$ 50, female = 0& LR \\
\hline
age $>$ 46, qnt\_income $>$ 2, age $\leq$ 50, female = 1& HR  \\
\hline
age $>$ 46, qnt\_income $>$ 2, age $>$ 50, age $\leq$ 58 & LR  \\
\hline
age $>$ 46, qnt\_income $>$ 2, age $>$ 50, age $>$ 58 & LR  \\
\bottomrule
\end{tabular}
\end{flushleft}
\begin{tablenotes}
\footnotesize
\textit{Note:} Policy rule for individuals assigned to HR: high rate, MR: medium rate and LR: low rate.
\end{tablenotes}
\end{table}

\begin{table}[ht]
\caption{Unconstrained sequentially optimal Policy Trees of depth-4+1}
\begin{flushleft}
\begin{tabular}{l|c}
\toprule
Splitting variables and values & Treatment  \\ \midrule
age $>$ 44, female = 0, qnt\_income $>$ 2, age $\leq$ 50& HR  \\ \hline
age $>$ 44, female = 0, qnt\_income $>$ 2, age $>$ 50& LR \\ \hline
age $\leq$ 44, age $\leq$ 33, age $\leq$ 28, female = 0, age $\leq$ 27& MR  \\ \hline
age $\leq$ 44, age $\leq$ 33, age $\leq$ 28, female = 0, age $>$ 27& HR \\ \hline
age $\leq$ 44, age $\leq$ 33, age $\leq$ 28, female = 1, age $\leq$ 25& HR  \\ \hline
age $\leq$ 44, age $\leq$ 33, age $\leq$ 28, female = 1, age $>$ 25& HR  \\ \hline
age $\leq$ 44, age $\leq$ 33, age $>$ 28, age $\leq$ 32, rural = 0& MR  \\ \hline
age $\leq$ 44, age $\leq$ 33, age $>$ 28, age $\leq$ 32, rural = 1& MR \\ \hline
age $\leq$ 44, age $\leq$ 33, age $>$ 28, age $>$ 32, female = 0& LR \\ \hline
age $\leq$ 44, age $\leq$ 33, age $>$ 28, age $>$ 32, female = 1& LR \\ \hline
age $\leq$ 44, age $>$ 33, edhi = 0, age $\leq$ 35, qnt\_income = 0& HR  \\ \hline
age $\leq$ 44, age $>$ 33, edhi = 0, age $\leq$ 35, qnt\_income $>$ 0 & LR \\ \hline
age $\leq$ 44, age $>$ 33, edhi = 0, age $>$ 35, qnt\_income $\leq$ 2& HR \\ \hline
age $\leq$ 44, age $>$ 33, edhi = 0, age $>$ 35, qnt\_income $>$ 2& HR\\ \hline
age $\leq$ 44, age $>$ 33, edhi = 1, age $\leq$ 37, qnt\_income = 0& MR \\ \hline
age $\leq$ 44, age $>$ 33, edhi = 1, age $\leq$ 37, qnt\_income $>$ 0& MR\\ \hline
age $\leq$ 44, age $>$ 33, edhi = 1, age $>$ 37, rural = 0& HR  \\ \hline
age $\leq$ 44, age $>$ 33, edhi = 1, age $>$ 37, rural = 1& HR  \\ \hline
age $>$ 44, female = 0, qnt\_income $\leq$ 2, age $\leq$ 53, qnt\_income = 0& HR \\ \hline
age $>$ 44, female = 0, qnt\_income $\leq$ 2, age $\leq$ 53, qnt\_income $>$ 0& MR\\ \hline
age $>$ 44, female = 0, qnt\_income $\leq$ 2, age $>$ 53, age $\leq$ 55& HR \\ \hline
age $>$ 44, female = 0, qnt\_income $\leq$ 2, age $>$ 53, age $>$ 55& HR \\ \hline
age $>$ 44, female  = 1 , age $\leq$ 53, age $\leq$ 50, age $\leq$ 47 & HR  \\ \hline
age $>$ 44, female  = 1 , age $\leq$ 53, age $\leq$ 50, age $>$ 47 & HR\\ \hline
age $>$ 44, female  = 1 , age $\leq$ 53, age $>$ 50, edhi = 0& HR  \\ \hline
age $>$ 44, female  = 1 , age $\leq$ 53, age $>$ 50, edhi = 1& MR \\ \hline
age $>$ 44, female  = 1 , age $>$ 53, edhi = 0, age $\leq$ 65& HR  \\ \hline
age $>$ 44, female  = 1 , age $>$ 53, edhi = 0, age $>$ 65& MR\\ \hline
age $>$ 44, female  = 1 , age $>$ 53, edhi = 1, age $\leq$ 58& LR \\ \hline
age $>$ 44, female = 1 , age $>$ 53, edhi = 1, age $>$ 58& LR  \\ \bottomrule
\end{tabular}
\end{flushleft}
\begin{tablenotes}
\footnotesize
\textit{Note:} Policy rule for individuals assigned to HR: high rate, MR: medium rate and LR: low rate.
\end{tablenotes}
\end{table}

\begin{table}[ht]
\caption{Unconstrained sequentially optimal Policy Trees of depth-2+2}
\begin{flushleft}    
\begin{tabular}{l|c}
\toprule
Splitting variables and values & Treatment allocation \\
\midrule
age $\leq$ 48, age $\leq$ 36, age $\leq$ 28, female = 0 & MR  \\
\hline
age $\leq$ 48, age $\leq$ 36, age $\leq$ 28, female $>$ 0 & HR \\
\hline
age $\leq$ 48, age $\leq$ 36, age $>$ 28, age $\leq$ 32 & MR \\
\hline
age $\leq$ 48, age $\leq$ 36, age $>$ 28, age $>$ 32 & LR \\
\hline
age $\leq$ 48, age $>$ 36, female = 0, age $\leq$ 46& HR \\
\hline
age $\leq$ 48, age $>$ 36, female = 0, age $>$ 46 & MR  \\
\hline
age $\leq$ 48, age $>$ 36, female $>$ 0, age $\leq$ 37 & MR \\
\hline
age $\leq$ 48, age $>$ 36, female $>$ 0, age $>$ 37 & HR  \\
\hline
age $>$ 48, qnt\_income $\leq$ 2, qnt\_income = 0, age $\leq$ 52 & HR  \\
\hline
age $>$ 48, qnt\_income $\leq$ 2, qnt\_income = 0, age $>$ 52 & HR  \\
\hline
age $>$ 48, qnt\_income $\leq$ 2, qnt\_income $>$ 0, age $\leq$ 53 & MR  \\
\hline
age $>$ 48, qnt\_income $\leq$ 2, qnt\_income$>$ 0, age $>$ 53 & HR \\
\hline
age $>$ 48, qnt\_income $>$ 2, age $\leq$ 51 & LR \\
\hline
age $>$ 48, qnt\_income $>$ 2, age $>$ 51 & LR\\
\bottomrule
\end{tabular}
\end{flushleft}
\begin{tablenotes}
\footnotesize
\textit{Note:} Policy rule for individuals assigned to HR: high rate, MR: medium rate and LR: low rate.
\end{tablenotes}
\end{table}

\begin{table}[ht]
\caption{Unconstrained sequentially optimal Policy Trees of depth-3+2}
\begin{flushleft}    
\begin{tabular}{l|c}
\toprule
Splitting variables and values & Treatment \\
& allocation\\
\midrule
age $\leq$ 44, age $\leq$ 28, female = 0, age $\leq$ 26, qnt\_income $\leq$ 1 & MR \\
\hline
age $\leq$ 44, age $\leq$ 28, female = 0, age $\leq$ 26, qnt\_income $>$ 1 & MR \\
\hline
age $\leq$ 44, age $\leq$ 28, female = 0, age $>$ 26, qnt\_income $\leq$ 1 & HR  \\
\hline
age $\leq$ 44, age $\leq$ 28, female = 0, age $>$ 26, qnt\_income $>$ 1 & MR \\
\hline
age $\leq$ 44, age $\leq$ 28, female = 1, qnt\_income $\leq$ 1, qnt\_income = 0 & HR \\
\hline
age $\leq$ 44, age $\leq$ 28, female = 1, qnt\_income $\leq$ 1, qnt\_income $>$ 0 & HR \\
\hline
age $\leq$ 44, age $\leq$ 28, female =1 , qnt\_income $>$ 1, qnt\_income $\leq$ 2.5 & HR \\
\hline
age $\leq$ 44, age $\leq$ 28, female $>$ 1, qnt\_income $>$ 1, qnt\_income $>$ 2.5 & HR \\
\hline
age $\leq$ 44, age $>$ 28, age $\leq$ 36, age $\leq$ 33, age $\leq$ 32 & MR  \\
\hline
age $\leq$ 44, age $>$ 28, age $\leq$ 36, age $\leq$ 33, age $>$ 32 & LR \\
\hline
age $\leq$ 44, age $>$ 28, age $\leq$ 36, age $>$ 33, edhi = 0 & LR  \\
\hline
age $\leq$ 44, age $>$ 28, age $\leq$ 36, age $>$ 33, edhi = 1 & MR \\
\hline
age $\leq$ 44, age $>$ 28, age $>$ 36, female = 0, qnt\_income $\leq$ 1 & HR  \\
\hline
age $\leq$ 44, age $>$ 28, age $>$ 36, female = 0, qnt\_income $>$ 1 & HR \\
\hline
age $\leq$ 44, age $>$ 28, age $>$ 36, female = 1 , qnt\_income $\leq$ 2 & HR \\
\hline
age $\leq$ 44, age $>$ 28, age $>$ 36, female = 1 , qnt\_income $>$ 2. & MR \\
\hline
age $>$ 44, qnt\_income $\leq$ 2, female = 0, age $\leq$ 55, qnt\_income = 0 & HR \\
\hline
age $>$ 44, qnt\_income $\leq$ 2, female = 0, age $\leq$ 55, qnt\_income $>$ 0 & MR \\
\hline
age $>$ 44, qnt\_income $\leq$ 2, female = 0, age $>$ 55, age $\leq$ 61 & HR\\
\hline
age $>$ 44, qnt\_income $\leq$ 2, female = 0, age $>$ 55, age $>$ 61 & HR \\
\hline
age $>$ 44, qnt\_income $\leq$ 2, female = 1, age $\leq$ 50, age $\leq$ 46 & HR\\
\hline
age $>$ 44, qnt\_income $\leq$ 2, female = 1 , age $\leq$ 50, age $>$ 46 & HR  \\
\hline
age $>$ 44, qnt\_income $\leq$ 2, female = 1 , age $>$ 50, age $\leq$ 53 & MR  \\
\hline
age $>$ 44, qnt\_income $\leq$ 2, female = 1 , age $>$ 50, age $>$ 53 & HR \\
\hline
age $>$ 44, qnt\_income $>$ 2, age $\leq$ 50, age $\leq$ 47 & HR \\
\hline
age $>$ 44, qnt\_income $>$ 2, age $\leq$ 5 age $>$ 47 & HR \\
\hline
age $>$ 44, qnt\_income $>$ 2, age $>$ 50, age $\leq$ 59 & LR \\
\hline
age $>$ 44, qnt\_income $>$ 2, age $>$ 50, age $>$ 59 & LR \\
\bottomrule
\end{tabular}
\end{flushleft}
\begin{tablenotes}
\footnotesize
\textit{Note:} Policy rule for individuals assigned to HR: high rate, MR: medium rate and LR: low rate.
\end{tablenotes}
\end{table}

\begin{table}[ht]
\footnotesize 
\caption{Unconstrained sequentially optimal Policy Trees of depth-4+2}
\begin{flushleft}    
\begin{tabular}{l|c}
\toprule
Splitting variables and values & Treatment \\
& allocation\\
\midrule
age $\leq$ 44, age $\leq$ 33, age $\leq$ 28, female = 0, age $\leq$ 26, qnt\_income $\leq$ 1 & MR \\
\hline
age $\leq$ 44, age $\leq$ 33, age $\leq$ 28, female = 0, age $\leq$ 26, qnt\_income $>$ 1 & MR \\
\hline
age $\leq$ 44, age $\leq$ 33, age $\leq$ 28, female = 0, age $>$ 26, qnt\_income $\leq$ 1 & HR \\
\hline
age $\leq$ 44, age $\leq$ 33, age $\leq$ 28, female = 0, age $>$ 26, qnt\_income $>$ 1 & MR\\
\hline
age $\leq$ 44, age $\leq$ 33, age $\leq$ 28, female $>$ 1, qnt\_income $\leq$ 1, qnt\_income = 0 & HR \\
\hline
age $\leq$ 44, age $\leq$ 33, age $\leq$ 28, female = 1, qnt\_income $\leq$ 1, qnt\_income $>$ 0 & HR\\ \hline
age $\leq$ 44, age $\leq$ 33, age $\leq$ 28, female = 1, qnt\_income $>$ 1, qnt\_income $\leq$ 2.5 & HR \\ \hline
age $\leq$ 44, age $\leq$ 33, age $\leq$ 28, female = 1, qnt\_income $>$ 1, qnt\_income $>$ 2.5  & HR \\ \hline
age $\leq$ 44, age $\leq$ 33, age $>$ 28, age $\leq$ 32, edhi = 0, age $\leq$ 30 & MR \\ \hline
age $\leq$ 44, age $\leq$ 33, age $>$ 28, age $\leq$ 32, edhi = 0, age $>$ 30 & HR\\ \hline
age $\leq$ 44, age $\leq$ 33, age $>$ 28, age $\leq$ 32, edhi = 1, qnt\_income = 0 & HR\\ \hline
age $\leq$ 44, age $\leq$ 33, age $>$ 28, age $\leq$ 32, edhi = 1, qnt\_income $>$ 0 & MR\\ \hline
age $\leq$ 44, age $\leq$ 33, age $>$ 28, age $>$ 32, female = 0 & LR\\ \hline
age $\leq$ 44, age $\leq$ 33, age $>$ 28, age $>$ 32, female $>$ 0 & LR \\ \hline
age $\leq$ 44, age $>$ 33, edhi = 0, age $\leq$ 35, qnt\_income = 0  & HR\\ \hline
age $\leq$ 44, age $>$ 33, edhi = 0, age $\leq$ 35, qnt\_income $>$ 0  & LR \\ \hline
age $\leq$ 44, age $>$ 33, edhi = 0, age $>$ 35, qnt\_income $\leq$ 2, age $\leq$ 36 & LR \\ \hline
age $\leq$ 44, age $>$ 33, edhi = 0, age $>$ 35, qnt\_income $\leq$ 2, age $>$ 36  & HR \\ \hline
age $\leq$ 44, age $>$ 33, edhi = 0, age $>$ 35, qnt\_income $>$ 2, age $\leq$ 39  & HR \\ \hline
age $\leq$ 44, age $>$ 33, edhi = 0, age $>$ 35, qnt\_income $>$ 2, age $>$ 39   & HR \\ \hline
age $\leq$ 44, age $>$ 33, edhi = 1, age $\leq$ 37, qnt\_income $\leq$ 2, qnt\_income $\leq$ 1  & MR\\ \hline
age $\leq$ 44, age $>$ 33, edhi = 1, age $\leq$ 37, qnt\_income $\leq$ 2, qnt\_income $>$ 1  & LR \\ \hline
age $\leq$ 44, age $>$ 33, edhi = 1, age $\leq$ 37, qnt\_income $>$ 2, female = 0  & MR\\ \hline
age $\leq$ 44, age $>$ 33, edhi = 1, age $\leq$ 37 qnt\_income $>$ 2, female = 1  & MR\\ \hline
age $\leq$ 44, age $>$ 33, edhi = 1, age $>$ 37, qnt\_income $\leq$ 2, qnt\_income $\leq$ 1 & HR \\ \hline
age $\leq$ 44 age $>$ 33, edhi = 1, age $>$ 37, qnt\_income $\leq$ 2, qnt\_income $\>$ 1  & HR \\
\hline
age $\leq$ 44, age $>$ 33, edhi = 1, age $>$ 37, qnt\_income $>$ 2, female = 0 & HR\\ \hline
age $\leq$ 44, age $>$ 33, edhi = 1, age $>$ 37, qnt\_income $>$ 2, female = 1 & MR \\ \hline
age $>$ 44, qnt\_income $\leq$ 1, age $\leq$ 52, female = 0, qnt\_income = 0 & HR \\
\hline
age $>$ 44, qnt\_income $\leq$ 1 age $\leq$ 52, female = 0, qnt\_income $>$ 0 & MR \\
\hline
age $>$ 44, qnt\_income $\leq$ 1, age $\leq$ 52, female = 1, age $\leq$ 47, qnt\_income = 0 & HR \\
\hline
age $>$ 44, qnt\_income $\leq$ 1, age $\leq$ 52, female = 1 , age $\leq$ 47, qnt\_income $>$ 0 & MR\\
\hline
age $>$ 44, qnt\_income $\leq$ 1, age $\leq$ 52, female = 1 , age $>$ 47, qnt\_income = 0  & MR\\
\hline
age $>$ 44, qnt\_income $\leq$ 1 age $\leq$ 52, female= 1, age $>$ 47, qnt\_income $>$ 0   & MR \\
\hline
age $>$ 44, qnt\_income $\leq$ 1, age $>$ 52, age $\leq$ 60, qnt\_income = 0, age $\leq$ 56   & HR\\
\hline
age $>$ 44, qnt\_income $\leq$ 1, age $>$ 52, age $\leq$ 60, qnt\_income = 0, age $>$ 56  & HR\\
\hline
age $>$ 44, qnt\_income $\leq$ 1, age $>$ 52, age $\leq$ 60, qnt\_income $>$ 0, age $\leq$ 55  & MR\\
\hline
age $>$ 44, qnt\_income $\leq$ 1, age $>$ 52, age $\leq$ 60, qnt\_income $>$ 0, age $>$ 55  & HR\\
\hline
age $> 44$, qnt\_income $\leq 1$, age $> 52$, age $> 60$, age $\leq 65$, age $\leq 62$   & HR\\ \hline
age $> 44$, qnt\_income $\leq 1$, age $> 52$, age $> 60$, age $\leq 65$, age $> 62$    & HR \\ \hline
age $> 44$, qnt\_income $\leq 1$, age $> 52$, age $> 60$, age $> 65$, age $\leq 68$ & HR  \\ \hline
age $> 44$, qnt\_income $\leq 1$, age $> 52.$, age $> 60$, age $> 65$, age $> 68$ & HR \\ \hline
age $> 44$, qnt\_income $> 1$, qnt\_income $\leq 2.5$, female = 0, age $\leq 55$, age $\leq 50$ & MR \\ \hline
age $> 44$, qnt\_income $> 1$, qnt\_income $\leq 2.5$, female = 0, age $\leq 55$, age $> 50$ & MR \\ \hline
age $> 44$, qnt\_income $> 1$, qnt\_income $\leq 2.5$, female = 0, age $> 55$, age $\leq 60$  & LR \\ \hline
age $> 44$, qnt\_income $> 1$, qnt\_income $\leq 2.5$, female = 0, age $> 55$, age $> 60$  & HR \\ \hline
age $> 44$, qnt\_income $> 1$, qnt\_income $\leq 2.5$, female = 1, age $\leq 53$, age $\leq 48$ & HR \\ \hline
age $> 44$, qnt\_income $> 1$, qnt\_income $\leq 2.5$, female = 1, age $\leq 53$, age $> 48$ & MR \\ \hline
age $> 44$, qnt\_income $> 1$, qnt\_income $\leq 2.5$, female = 1, age $> 53$, age $\leq 62$ & HR  \\ \hline
age $> 44$, qnt\_income $> 1$, qnt\_income $\leq 2.5$, female = 1, age $> 53$, age $> 62$ & HR  \\ \hline
age $> 44$, qnt\_income $> 1$, qnt\_income $> 2.5$, age $\leq 50$, age $\leq 46$, female = 0  & HR \\ \hline
age $> 44$, qnt\_income $> 1$, qnt\_income $> 2.5$, age $\leq 50$, age $\leq 46$, female = 1 & LR \\ \hline
age $> 44$, qnt\_income $> 1$, qnt\_income $> 2.5$, age $\leq 50$, age $> 46$, female = 0 & LR\\ \hline
age $> 44$, qnt\_income $> 1$, qnt\_income $> 2.5$, age $\leq 50$, age $> 46$, female $> 0$ & HR\\ \hline
age $> 44$, qnt\_income $> 1$, qnt\_income $> 2.5$, age $> 50$, age $\leq 55$, female = 0 & LR\\ \hline
age $> 44$, qnt\_income $> 1$, qnt\_income $> 2.5$, age $> 50$, age $\leq 55$, female = 1 & LR \\ \hline
age $> 44$, qnt\_income $> 1$, qnt\_income $> 2.5$, age $> 50$, age $> 55$, female = 0 & LR\\ \hline
age $> 44$, qnt\_income $> 1$, qnt\_income $> 2.5$, age $> 50$, age $> 55$, female = 1 & LR \\ \bottomrule
\end{tabular}
\end{flushleft}
\begin{tablenotes}
\footnotesize
\textit{Note:} Policy rule for individuals assigned to HR: high rate, MR: medium rate and LR: low rate.
\end{tablenotes}
\end{table}